\documentclass[journal abbreviation, manuscript]{copernicus}

\title{Multivariate phase-space interpretation of ensemble forecasts: A comparison of physical and AI models}

\Author[1][martin.bonte@meteo.be]{Martin}{Bonte}
\Author[2]{Stéphane}{Vannitsem}

\affil[1]{Royal Meteorological Institute of Belgium, Brussels, Belgium}
\affil[2]{School of Physical and Mathematical Sciences \& The Asian School of the Environment, Nanyang Technological University, Singapore}

\begin{document}

\nolinenumbers

\firstpage{1}

\runningtitle{Phase-space interpretation of ensemble forecasts}
\runningauthor{M. Bonte and S. Vannitsem}

\maketitle

\begin{abstract}
Ensemble forecasts from the  ECMWF's Integrated Forecasting System (IFS) and the Artificial Intelligence Forecasting System (AIFS-ENS) are compared using a multi-dimensional evaluation framework.
Within this framework, the uncertainty is interpreted as the linear subspace spanned by the ensemble members in phase space, from which various geometric properties are extracted. As revealed by the analysis, the error does not entirely lie in this subspace, but is reasonably well calibrated in all directions. Moreover, the framework allows to quantify that this subspace critically depends on the number of ensemble members, suggesting that the current setups only partially represent the uncertainty subspace.
This phase-space interpretation furthermore allows a direct comparison of the error-representation patterns of AIFS-ENS and IFS by computing the principal angles between the subspaces: IFS and AIFS ensemble subspaces do not entirely overlap in phase space, even though they were shown to have similar abilities to represent the amplitude and the structure of the error.

\end{abstract}


\section{Introduction}

Ensemble forecasts are often evaluated using univariate probabilistic scores. For instance, to verify spatially extended variables, the same metrics are often computed at each location separately and then averaged over all locations. This approach implicitly assumes that ensemble members at different locations are uncorrelated, ignoring any spatial patterns among them. Examples of such scores include the rank histogram, the Continuous Ranked Probability Score (CRPS) and the spread-RMSE ratio.

Yet, techniques for generating ensemble forecasts were developed from dynamical systems considerations, where variables live in multi-dimensional spaces. Singular vectors were developed to construct initial perturbations maximizing the forecast error after some time (see \cite{Kalnay2019}, \cite{Palmer2019} and \cite{Buizza2019} for historical perspectives). These perturbations are vectors in phase space with specific structures and, from general arguments about the error growth in dynamical systems, the error is expected to progressively align along the associated propagated vectors \citep{Farrell1990,Molteni1993,Kalnay1998,Leutbecher2008}. Therefore, it is quite natural to use conceptually similar tools to interpret and evaluate ensembles.

Moreover, meaningful verification techniques are needed more than ever as artificial intelligence (AI) models develop. Indeed, training such models often optimizes global statistics of uni-dimensional metrics, which are also often used to evaluate these models. Detailed assessments of these models' forecasts are still needed for operational use \citep{Benbouallegue2024, Radford2025, Brocker2026}.

Several techniques have already been proposed to verify ensembles of multi-dimensional forecasts. Building on rank histograms for uni-dimensional variables, \cite{Gneiting2008} proposed to use a so-called "pre-rank function" to rank the multi-dimensional observations and ensemble members, and to build rank histograms from the output of this function. A special case of this technique is, for instance, the rank histograms of the lengths of the Minimum Spanning Tree (MST) defined by the ensemble members in phase space \citep{Wilks2004} and \cite{Smith2004}. Several choices of pre-rank functions are evaluated in \cite{Wilks2017} and \cite{Thorarinsdottir2018}. Another possibility is to use proper scoring rules: \cite{Gneiting2007} proposed the energy score to generalize the CRPS to multi-dimensional variables and \cite{Scheuerer2015} introduced the variogram score to evaluate the dependence structure of the forecasts.

This work contributes to building an interpretation and evaluation framework for ensemble forecasts as geometric objects in phase space, inspired by the description of error growth in dynamical systems. An ensemble is interpreted as the linear subspace spanned by its ensemble members in phase space, and it is considered to perfectly represent the error if it is calibrated in all directions. This is evaluated using geometric quantities in phase space, following ideas from \cite{Buizza1998, Wei2003, Szunyogh2005, Satterfield2010, Satterfield2011, Leutbecher2014} and \cite{Bonte2026}. These techniques use the decomposition of the ensemble into Empirical Orthogonal Functions (EOFs), providing an easy way to understand the structure of the ensemble. While several tools used in this work are not new in ensemble verification contexts, their use on an AI model's forecasts is new to the best of the authors' knowledge.

The ensembles evaluated in this work are those of the AI model AIFS-CRPS \citep{Lang2026} and of the NWP model IFS. Surrogate ensembles are also constructed in a way that the perturbations in these ensembles have the same power spectra as those in AIFS-ENS and IFS ensembles. Comparing the computed metrics of the original ensembles with those of the corresponding surrogate ensembles serves as a meta-verification technique because it reveals how much of the original ensembles' scores is due to real skill rather than to the power spectra of the perturbations.

Section \ref{forecasts_analyses_sec} presents the data used in this work. The evaluation methods are presented in Sec. \ref{evaluation_methods_sec}: the ensemble covariance matrix, the ensemble subspaces are introduced in Sec. \ref{covariance_eigendecomposition_sec}, $\cos\gamma(e, p)$, its interpretation and the link with other metrics are presented in Sec. \ref{cos_gamma_PECA_sec}, the technique to evaluate ensemble calibration in all directions is detailed in Sec. \ref{projection_p_u_i_relative_variances_sec}, Sec. \ref{principal_angles_def_sec} exposes how to use principal angles to compare two linear subspaces, and the surrogate ensembles are introduced in Sec. \ref{surrogates_sec}. The results are presented in Sec. \ref{results_sec}, where the size of the ensemble to estimate the covariance matrix is first evaluated. Section \ref{discussion_conclusion_sec} summarizes the results and discusses their overall interpretation, together with potential future work.

\section{Forecasts and verifying analyses used in this work}

\label{forecasts_analyses_sec}

Ensembles of global forecasts of the geopotential at 500 hPa from IFS (CY49R1 cycle) and from AIFS-ENS (v1 version) are analyzed. 100 forecasts are selected randomly between July and November 2025. AIFS-ENS is ECMWF's ensemble forecasting system based on AIFS-CRPS \citep{Lang2026}. Details on AIFS-ENS can be found on ECMWF's website\footnote{https://www.ecmwf.int/en/forecasts/datasets/set-x}. Details on IFS are in the documentation \citep{IFSCY49R1}.

The forecasts were downloaded on a regular 0.25x0.25 grid, but the computation of scalar products between vectors in phase space is done in spectral space (using spherical harmonics) to avoid contributions from oversampled points near the poles. 

AIFS-CRPS is trained with the almost-fair CRPS, and each member is obtained from the same initial condition during training. During inference, AIFS-ENS members are generated using the operational IFS initial conditions, interpolated from their native O1280 resolution (approximately 0.1°) to the AIFS-CRPS resolution (N320, approximately 0.25°).

Details of the IFS initial conditions are in \cite{Lang2019} and \cite{IFSCY49R1}. They are built using the method proposed in \cite{Buizza2008}, where the initial conditions are obtained as
\begin{align}
    x_i = x_a + (eda_i - \bar{eda}) + sv_i,
    \label{IFS_perturbations_eq}
\end{align}
with $x_a$ being the deterministic analysis, $eda_i$ a member of the Ensemble Data Assimilation (EDA), $\bar{eda}$ the mean of EDA members, and $sv_k$ a perturbation based on singular vectors. These perturbations are linear combinations of singular vectors, with coefficients randomly drawn from a Gaussian distribution with zero mean \citep{Leutbecher2008}, while the variance is tuned to yield calibrated ensembles \citep{IFSCY49R1}.

The verifying analyses are the geopotential fields of ERA5 \citep{Hersbach2023} at 500 hPa for which the AIFS-ENS and IFS forecasts are valid, at the same resolution.

\section{Evaluation methods}

\label{evaluation_methods_sec}

The ensemble members of the AIFS-ENS and IFS forecasts are denoted $x_i$ for $i = 1, ..., N$, with $N = 50$. The time dependence is omitted for the sake of clarity. The verifying analysis is $y$, while the ensemble mean, the error, and the residual vectors are, respectively,
\begin{align}
    \bar{x} &= \frac{1}{N}\sum_i x_i,\\
    e &= y - \bar{x},\\
    v_i &= x_i - \bar{x}.
\end{align}

\subsection{Ensemble covariance matrix and eigen-decomposition}

\label{covariance_eigendecomposition_sec}

The covariance matrix of the residual vectors $v_i$ in phase space is used in this work to characterize phase space directions in which the ensembles are extended as well as the amplitudes of these directions. If the component $\alpha$ of the $i$th residual vector of the ensemble is denoted as $v_{i, \alpha}$, the ensemble covariance matrix is defined as
\begin{align}
    C_{\alpha\beta} = \frac{1}{N-1}\sum_{i = 1}^N v_{i,\alpha}v_{i, \beta}.
\end{align}
Since the $v_i$ satisfy $\sum_iv_i = 0$ (and this is generically the only linear relationship between the $v_i$), the ensemble covariance matrix has rank $N-1$. The eigenvectors of this covariance matrix are denoted $u_i$ for $i = 1, ..., N-1$ (and the component of $u_i$ in direction $\alpha$ is $u_{i, \alpha}$). These vectors are orthonormal: $u_i \cdot u_j = \sum_\alpha u_{i, \alpha}u_{j, \alpha} = \delta_{ij}$.

When the ensemble is considered as a cloud of points in phase space, the covariance matrix describes how each phase-space dimension co-varies with other directions among that set of points. An eigenvalue of the covariance matrix is the variance of the ensemble in the direction given by the associated eigenvectors. Computing the covariance matrix and working with its eigenvectors is exactly the same as decomposing the ensemble in Empirical Orthogonal Functions (EOFs, see \cite{Hannachi2021} for a review) and keeping all the modes.

The covariance matrix is huge for the high-dimensional phase spaces of weather forecasting, so computing it can be very costly. In practice, however, a shortcut is possible using the link between the Singular Value Decomposition (SVD) of a matrix $V$ and the diagonalization of $VV^T$. Let $V$ be the matrix whose columns are the $v_i$. The SVD of $V$ is
\begin{align}
V = U\Sigma W^T
\label{SVD_V}
\end{align}
with $U^TU = I$, $W^TW = I$ and $\Sigma$ a diagonal matrix whose elements are $\sigma_i$. The covariance matrix is then
\begin{align}
    C = \frac{1}{N-1}VV^T = U\frac{\Sigma^2}{N-1}U^T.
\end{align}
Since the matrix $\frac{\Sigma^2}{N-1}$ is diagonal and $U$ is orthonormal, the second equality is actually a diagonalization of $C$: the eigenvectors $u_i$ are the columns of $U$ and the eigenvalues are $\lambda_i = \frac{\sigma^2_i}{N-1}$.

The equality Eq. \eqref{SVD_V} also shows that the vectors $v_i$ can be written as linear combinations of $u_i$, and vice versa because Eq. \eqref{SVD_V} can be inverted to $U = VW\Sigma^{-1}$, so that the $v_i$ and the $u_i$ actually span the same linear subspace in phase space. This is essentially because the eigenvectors span the subspace in which the ensemble lives, i.e., the subspace in which it has some variance. This subspace will be called the "ensemble subspace" in the following.

This subspace can be described by both the $v_i$ and the $u_i$, but the $v_i$ are redundant because of $\sum_i v_i = 0$. The $u_i$ are, in addition, an orthonormal basis for that subspace, which makes it easier to compute the scalar product or the projection of a vector onto the subspace. The $u_i$ are also eigenvectors of the covariance matrix, and they can be interpreted as the ensemble's principal components: the associated eigenvalues order them by the variance they explain.

As discussed in detail below (Sec. \ref{projection_p_u_i_relative_variances_sec}), if the ensemble is expected to be calibrated as in \cite{Fortin2014} for each direction, the amplitude of the ensemble in each direction, as measured by the variance in that direction, should be a measure of the average error in that direction. This implies that the error should lie entirely in the ensemble subspace. A measure of this latter property is presented in the next Section.
The eigenvectors of the covariance matrix of an ensemble of forecasts have already been computed, for example in \cite{Molteni1999}, but also by \cite{Szunyogh2005}, \cite{Satterfield2010} and \cite{Satterfield2011} to compute the Explained Variance (see Sec. \ref{cos_gamma_sec}), and in \cite{Zheng2019} to evaluate operational ensemble prediction systems. The covariance matrix of an ensemble is also often used in data assimilation contexts \cite{Carrassi2018}.

Examples of eigenvectors for an AIFS-ENS ensemble of forecasts and for the corresponding IFS ensemble are shown in the Supplementary material.

\subsection{Projection on ensemble subspace}

\label{cos_gamma_sec}

When interpreting an ensemble through its linear subspace in phase space, any vector in this subspace is considered as a perturbation described by the ensemble. Conversely, if a vector has a component orthogonal to the subspace, this component is a perturbation not included in the ensemble. Therefore, if the ensemble completely represents the error, the error vector should lie entirely within the ensemble subspace. As shown in Sec. \ref{projection_p_u_i_relative_variances_sec}, this is also a necessary condition for the ensemble to be calibrated in all directions.

The cosine of the error vector with its projection onto the ensemble subspace measures how much of the error is captured in the subspace. This projection is
\begin{align}
    p = \sum_i (e\cdot u_i) u_i.
    \label{p_def}
\end{align}
The cosine of the angle between $e$ and $p$ is
\begin{align}
    \cos\gamma(e, p) = \frac{e\cdot p}{|e||p|}.
\end{align}
and is a measure of the difference between the two vectors, independently of their norms. Since
\begin{align}
    e\cdot p = \sum_i(e\cdot u_i)^2 = |p|^2,
\end{align}
$\cos\gamma(e, p)$ is also equal to
\begin{align}
    \cos\gamma(e, p) = \frac{|p|}{|e|},
\end{align}
so that $\cos\gamma(e, p)$ can be interpreted either as the cosine of the angle between $e$ and $p$, or as the fraction of $e$ explained by its projection on the ensemble subspace. By construction, $\cos\gamma(e, p)$ is smaller than $1$, and it is $1$ if $p$ can be completely projected onto the ensemble subspace. Conversely, $\cos\gamma(e, p) = 1$ does not mean that the ensemble is perfect because there might be "useless" directions in the ensemble, in the sense that the error has actually no chance to have a projection along them. These directions would be directions in which no error can grow (singular values smaller than $1$).

This quantity was introduced in various places in the literature, seemingly without always being identified as the same. \cite{Buizza1998} used it under the name of Projection Index to quantify the fraction of error explained by the first singular vectors. \cite{Wei2003} introduced the Perturbation versus Error Correlation Analysis (PECA) score, which is equal to $\cos\gamma(e, p)$ if the control run is replaced by the ensemble mean (see Appendix \ref{cos_gamma_PECA_sec} for details). \cite{Szunyogh2005,Satterfield2010} and \cite{Satterfield2011} introduced $\cos\gamma(e, p)$ as the Explained Variance. \cite{Uboldi2015} also used that metric to evaluate the ability of breeding vectors to capture the error growth in convective situations. The name $\cos\gamma(e, p)$ is kept in this work because of the geometric interpretation it conveys.

\subsection{Projection on eigenvectors and relative variances in ensembles}

\label{projection_p_u_i_relative_variances_sec}

Calibration for an ensemble usually means that the total error amplitude is, on average, equal to the total variance \citep{Fortin2014}:
\begin{align}
    |e|^2 = \sum_i \lambda_i.
    \label{calibration_def}
\end{align}
The right-hand side is indeed equal to $\frac{1}{N-1}\sum v_i^2$ using Eq. \eqref{SVD_V} and the identification $\lambda_i = \frac{\sigma^2_i}{N-1}$. To go a step further, the above equality is decomposed and verified along all the components of the ensemble. Explicitly, the ensemble is calibrated in the direction of $u_i$ if, on average,
\begin{align}
    (e\cdot u_i)^2 = \lambda_i.
    \label{calibration_in_each_direction_def}
\end{align}
One can show that this is equivalent to stating that, for any direction $\alpha$ in phase space, the variance of the $v_i$ along that direction (given by the variance of $v_i \cdot \alpha$) has to be equal to the squared component of the error along that direction.

Ensembles were evaluated exactly with this idea by \cite{Molteni1999} and the equivalent property for singular vectors was tested in \cite{Leutbecher2014}. \cite{Satterfield2010} and \cite{Satterfield2011} evaluated the same aspect but considered the mean of the ratio between $(e\cdot u_i)$ and $\lambda_i$ (instead of comparing the mean of each side of Eq. \eqref{calibration_in_each_direction_def}).

However, as pointed out in \cite{Leutbecher2014}, Eq. \eqref{calibration_in_each_direction_def} cannot hold if Eq. \eqref{calibration_def} holds and, at the same time, the error is not completely contained in the ensemble subspace. To see this, the error is decomposed into its parallel and perpendicular components to the ensemble subspace as
\begin{align}
    e = p + e^\perp
\end{align}
where $p$ in Eq. \eqref{p_def} is the parallel component. The squared norm of $|e|^2$ is
\begin{align}
    |e|^2 &= |p|^2 + |e^\perp|^2\\
    & = \sum_i(e\cdot u_i)^2 + |e^\perp|^2.
    \label{decomposition_error_eq}
\end{align}
Together with Eq. \eqref{calibration_def}, this implies that
\begin{align}
    \sum_i(e\cdot u_i)^2 + |e^\perp|^2 = \sum_i\lambda_i.
\end{align}
Obviously, the equality between $(e\cdot u_i)^2$ and $\lambda_i$ can only hold if $e^\perp = 0$, which is equivalent to $\cos\gamma(e, p) = 1$. As shown below, this is not the case, so it is not possible to have $(e\cdot u_i)^2 \approx \lambda_i$. This is why \cite{Leutbecher2014} tuned the singular values with a global factor, so that the first singular values match the error projections in those directions. 

One way to fix this problem is to compare the projection of $e$ onto $u_i$ relative to the projection onto all $u_i$ on one side, with $\lambda_i$ relative to all $\lambda_i$ on the other side. Explicitly, the equality to check in a statistical sense becomes
\begin{align}
    \frac{(e\cdot u_i)^2}{\sum_j (e\cdot u_j)^2} = \frac{\lambda_i}{\sum_j\lambda_j}.
\end{align}
The left-hand side is also equal to $\frac{(p\cdot u_i)^2}{|p|^2}$, which is in turn the square of the cosine between $p$ and $u_i$ (since $u_i$ is normalized). The above equality is thus
\begin{align}
    \cos^2\gamma(p, u_i) = \frac{\lambda_i}{\sum_j\lambda_j}.
\end{align}
This fixes the problem because the sum over $i$ of each side of this equality is $1$, allowing the possibility that this equality holds for each $i$.

\subsection{Principal angles}

\label{principal_angles_def_sec}

A way to compare two linear subspaces is needed in order to evaluate the convergence of the ensemble covariance matrices presented in Sec. \ref{ensemble_stability_sec}, and also to compare AIFS-ENS and IFS ensembles (Sec. \ref{angles_sec}). Principal angles \citep{Bjorck1973} naturally compare linear subspaces. They are used in numerical analysis to quantify the difference between the eigenspace of a matrix and that of a perturbed version of the matrix \citep{Stewart1990}. Principal angles are also used in Canonical Correlation Analysis, where their cosines are exactly the canonical correlation coefficients between two sets of variables \citep{Mardia2024}.

The principal angles between the subspaces $A$ and $B$ are defined recursively as follows. The first principal angle is defined as
\begin{align}
    \cos\theta_1 = \textrm{argmax } a\cdot b \quad \textrm{ for }
    \left\{\begin{array}{cc}
            a \in A,& |a| = 1\\
            b \in B,& |b| = 1
           \end{array}
    \right.
    \label{principal_angle_def}
\end{align}
It is the smallest angle between all pairs of unit vectors in the hyperplanes. Once $\theta_1$ has been found for some vectors $a^*$ and $b^*$, $\theta_2$ is defined in the same way, with the additional constraint that $a$ and $b$ have to be orthogonal to $a^*$ and $b^*$, etc. In practice, if $a_\alpha$ and $b_\beta$ are orthonormal bases of $A$ and $B$ respectively, the $\cos\theta_i$ can be computed as the singular values of the matrix whose elements are given by the scalar products $a_\alpha\cdot b_\beta$ (this matrix is of size $\textrm{dim}A\times \textrm{dim}B$). The vectors $a^*$ and $b^*$ associated with the first principal angle are the directions in each subspace in which $A$ and $B$ ensembles agree the most, and $\cos\theta_1$ is a measure of the agreement in this direction. Other principal angles and their associated vectors have similar interpretations in orthogonal directions. The number of principal angles is equal to the minimum value between $\textrm{dim}A$ and $\textrm{dim}B$. A more complete introduction can be found in \cite{Golub2013}.

Note that principal angles used on ensembles compare them as linear subspaces, i.e., they are seen as subspaces $A$ and $B$ where any linear combination of the basis vectors is allowed to maximize Eq. \ref{principal_angle_def}. This does not take into account the eigenvalues of the covariance matrix or any other ordering between the vectors of the ensembles if there is one (an ensemble of singular vectors is, for example, ordered by the singular values).

\subsection{Surrogate ensembles}

\label{surrogates_sec}

The metrics in this work are geometric quantities in phase space, whose dimension equals the number of independent spherical harmonics at the field resolution\footnote{The maximum order for the spherical harmonics expansion is $l_{MAX} = 359$ with the resolution of our data, and there are $2l + 1$ spherical harmonics for each $l \in \{0, ..., l_{MAX}\}$, so $D = \sum_{l=0}^{l_{MAX}}(2l+1) = (l_{MAX} + 1)^2 = 129600$} ($D = 129600$). In contrast, the dimension of the ensemble subspaces is $49$, which is small by comparison. Two vectors chosen randomly in phase space are almost always orthogonal, and the projection of one random vector on one randomly chosen subspace is almost always zero. Because of that, one would, for example, expect that $\cos\gamma(e, p)$ being significantly different from zero denotes a real skill of the ensemble.

\begin{figure}
\centering
\hspace*{-0cm}
\includegraphics[scale=0.5]{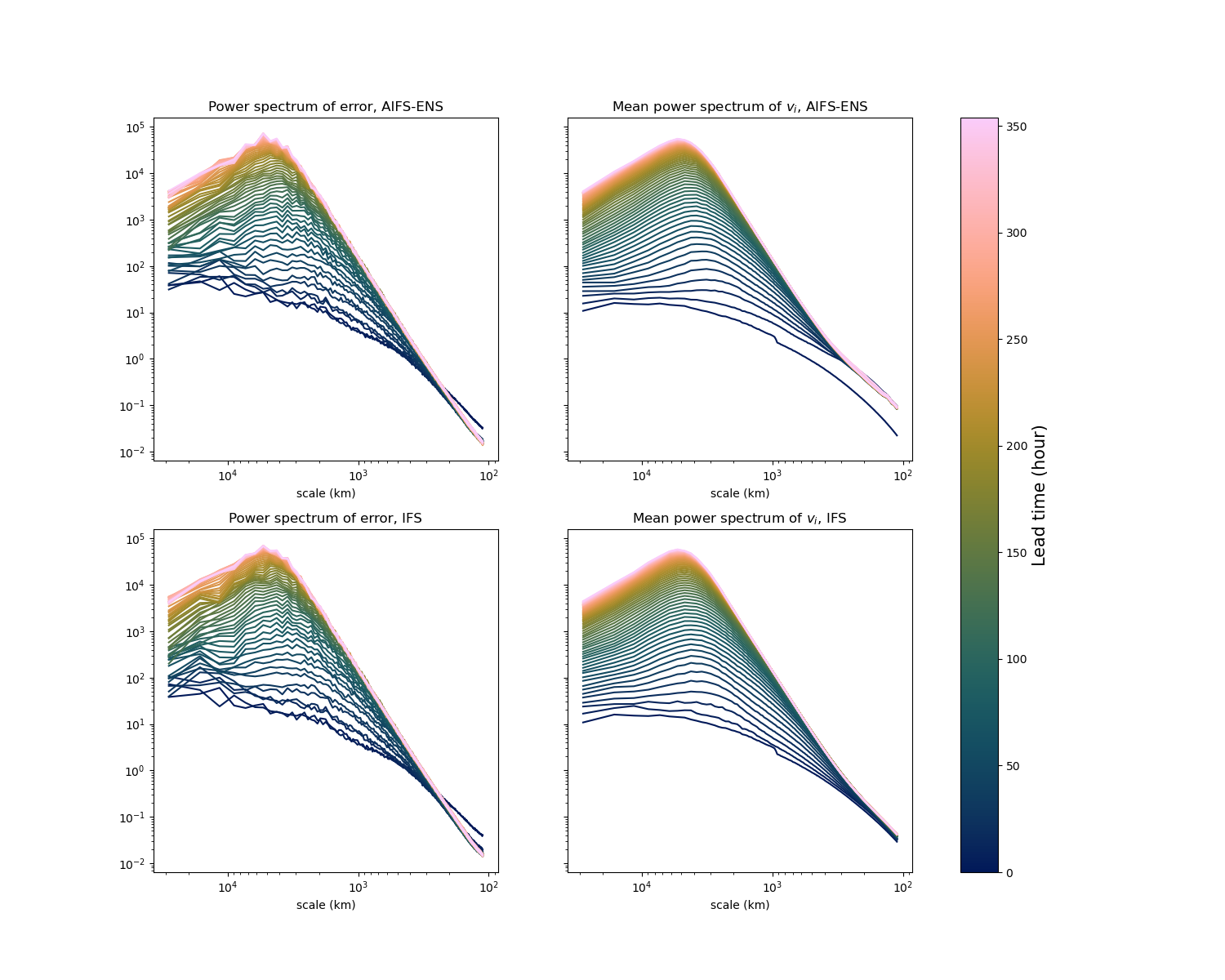}
\caption{Power spectrum of the error and of the $v_i$, for AIFS-ENS and IFS and for different lead times.}
\label{spectral_error_variance_fig}
\end{figure}

The phase space is, however, structured by the shape of the power spectra of the fields. Figure \ref{spectral_error_variance_fig} displays the power spectrum of the error $e$ and the mean power spectrum of the $v_i$ (which is also the total variance in the ensemble decomposed in scales, up to a factor of $\frac{N}{N-1}$), for AIFS-ENS and IFS. The power spectra were averaged over 100 forecasts. \cite{Rodwell2025} recently analyzed the power spectra of the variance and of the error in IFS ensembles and the ensembles of a prototype version of AIFS-CRPS.

Since spherical harmonics $Y_{lm}$ are orthogonal, they allow to represent any field on the sphere and can be used as a basis in phase space to represent a vector. If $a$ is a vector with a power spectrum as in Fig. \ref{spectral_error_variance_fig}, its coefficients $a_{lm}$ of the spherical expansion are such that $a_{lm}$ are generally smaller for larger values of $l$ (i.e. smaller spatial scales). There are several orders of magnitude between the values of the coefficients for different values of $l$, especially at the end of the forecast.

In the basis of spherical harmonics, the scalar product between two vectors $a$ and $b$ is
\begin{align}
    a\cdot b = \sum_{l = 0}^{L_{MAX}} \sum_{l = -m}^m a_{lm} b_{lm}^*.
    \label{scalar_product_SH_eq}
\end{align}
If $a$ and $b$ are two vectors with power spectra as in Fig. \ref{spectral_error_variance_fig}, the terms for large $l$ in Eq. \eqref{scalar_product_SH_eq} are completely negligible with respect to those for smaller values of $l$. At the end of the forecast, the scales having the largest contribution to the scalar product are those for which the power spectra peak (around 4000-5000km), the typical synoptic scale maximum related to the baroclinic instability (see e.g. \cite{Vannitsem1997,Reynolds1999,Corazza2003,Schubert2015}).

The large difference of magnitude between the coefficients makes coefficients of small scales irrelevant for the scalar product in \ref{scalar_product_SH_eq}, as if the corresponding dimensions in phase space did not exist. From the point of view of that scalar product, the effective dimension of the phase space is the number of non-negligible terms in Eq. \eqref{scalar_product_SH_eq}.

 Because of such a structure in phase space, random vectors with the same power spectrum as $e$ and $v_i$ have a high chance of having non-zero scalar products or of having a non-zero projection on each other at the end of the forecast. Because of this, $\cos\gamma(e, p)$ and $\cos\gamma(p, u_i)$ might be nonzero even if the $v_i$ have completely random orientations in phase space. Surrogate ensembles are therefore generated in this work from the AIFS-ENS and IFS ensembles in order to quantify how much the computed values of the metrics are really due to the skill of these models to produce $v_i$ with relevant orientations. AIFS-ENS and IFS ensembles are called "original ensembles" in the following if they need to be distinguished from their surrogates.

The surrogates are constructed member by member as $\bar{x} + w_i$, where $w_i$ is a statistically isotropic random field on the sphere with the same power spectrum as the $v_i$ of the corresponding member in the original ensemble. The coefficients $a_{lm}$ of spherical harmonics in the spherical expansion of $w_i$ are drawn from a normal distribution $\mathcal{N}(0, \rho_l)$, where $\rho_l^2$ is the power spectrum of $v_i$ at order $l$. When generated in this way, $w_i$ and $v_i$ do not have exactly the same power spectrum, but the $a_{lm}$ for a given $l$ can be rescaled after drawing them in order to have exactly $\frac{1}{2l + 1}\sum_{-m}^m |a_{lm}|^2 = \rho_l^2$. The results of this work did not change much when this exact matching was enforced, so the results presented here are obtained without this constraint. Statistically isotropic random fields are used in cosmology to test the Gaussianity of the Cosmic Microwave Background \citep{Dodelson2020}.

The surrogates used in this work differ slightly from the SPEC surrogates used in \cite{Bonte2026}. This is because constructing SPEC surrogates was found to alter the ensemble mean and affect the error. It is important that the ensemble mean does not change too much in constructing the surrogates, since the purpose of these surrogates is only to assess the effects of the power spectrum of $v_i$, which are vectors defined with respect to the ensemble mean.

\section{Results}

\label{results_sec}

\subsection{Convergence of the ensemble covariance matrix and of ensemble subspaces with the number of ensemble members}

\label{ensemble_stability_sec}

The covariance matrices of the ensembles are computed with a limited number of members, so assessing whether there are enough members to estimate these matrices is an important evaluation step. This amounts to testing whether the ensemble covariance matrix converges to the covariance matrix of the underlying distribution generating the ensemble.

To do so, three smaller ensembles are constructed from AIFS-ENS and IFS ensembles by randomly selecting 5, 10, and 20 members. The dependence of the eigenvalues and eigenvectors of the ensemble covariance matrix is then analyzed as a function of the number of members in these ensembles. If the ensembles contain enough members, the eigenvalues and eigenvectors of the covariance matrix should reach stable values for the largest ensembles.

\begin{figure}
\centering
\hspace*{-2cm}
\includegraphics[scale=0.55]{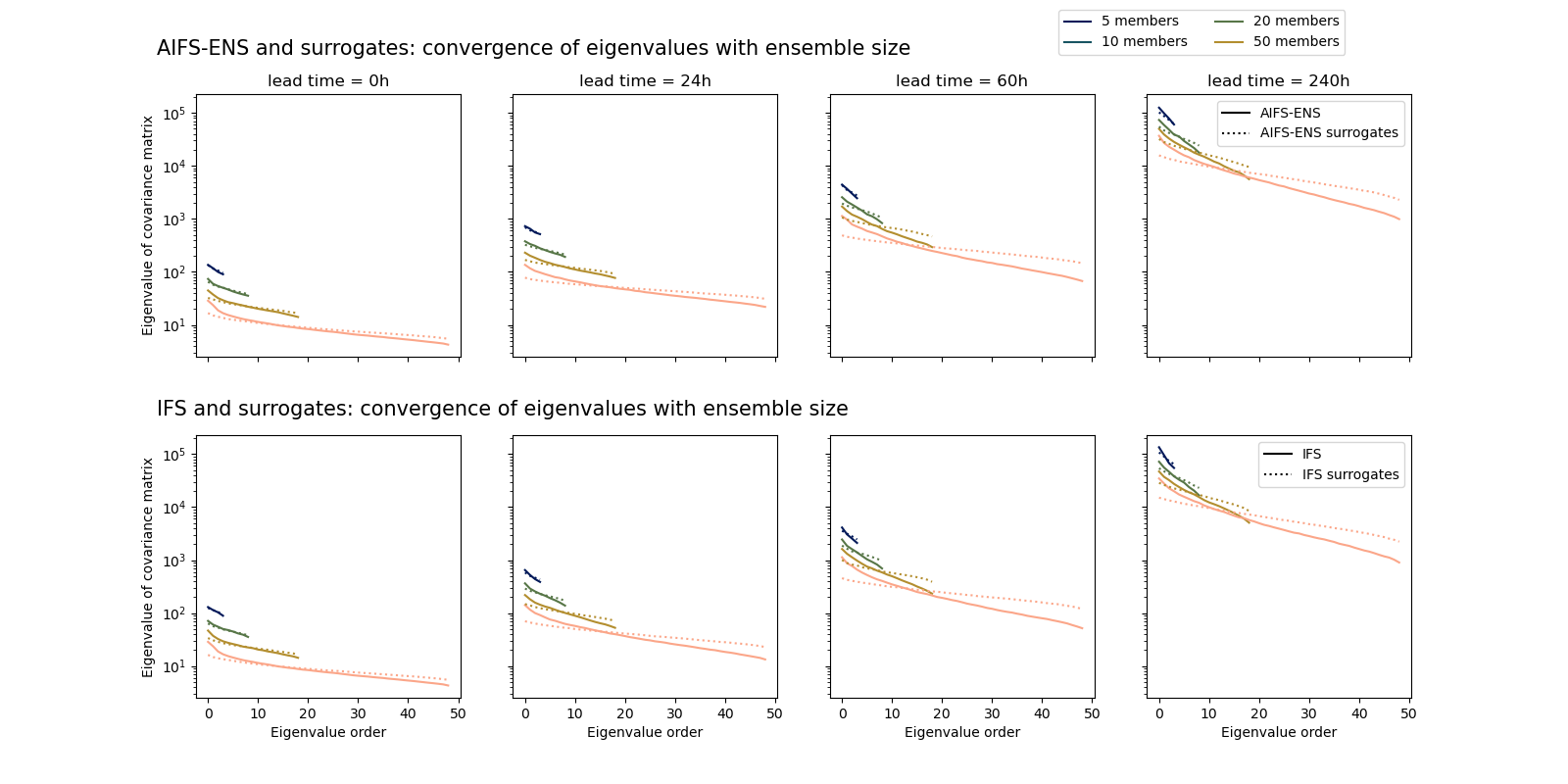}
\caption{Eigenvalues convergence for AIFS-ENS and surrogates (top row) and for IFS and IFS surrogates (bottom row), for different lead times. Note the logarithmic scale on the vertical axis.}
\label{eigenvalues_convergence_fig}
\end{figure}

Note that, since the subspace spanned by the $v_i$ is the ensemble subspace, an ensemble extracted from a larger ensemble always has its ensemble subspace included in the ensemble subspace of the larger ensemble. This supposes that the estimations of the ensemble mean in both ensembles are close enough; otherwise, the $v_i$ of corresponding members are different in the two ensembles.

\subsubsection*{Convergence of eigenvalues}

Figure \ref{eigenvalues_convergence_fig} shows the eigenvalue spectrum of the covariance matrix for different lead times for AIFS-ENS and IFS and their surrogate ensembles. The eigenvalues were averaged over 100 different forecasts. There are more eigenvalues in ensembles with more members since there are always fewer members than the dimension of the phase space, so only the accuracy of the estimates of the first eigenvalues can be assessed.

The eigenvalues generally decrease with the number of ensemble members because no ensemble members completely point in the same direction in phase space. Since the eigenvalues represent the variances in the directions of the eigenvectors, they are essentially equal to the size of the residual vectors $v_i$ for very small ensembles (because all $v_i$ are very likely orthogonal in that case). When a member is added, its $v_i$ vector is again likely orthogonal to other $v_i$, so that the addition of this member adds variance in another direction and decreases a little the variance in other directions. When the underlying distribution generating the ensemble starts to be well-sampled, additional $v_i$ have non-zero projections on previous $v_i$, and they will instead contribute to the variances of existing dimensions.

For all lead times, the eigenvalues are not stable with the number of ensemble members. Instead, they decrease regularly on the logarithmic scale in which the data is presented in Fig. \ref{eigenvalues_convergence_fig}. This shows that the ensembles do not have enough ensemble members to reliably estimate the covariance matrix of the underlying distribution. Note that the eigenvalues for the surrogate ensembles follow a similar pattern, except that the spectra are flatter, showing that the surrogate ensembles are more isotropic in phase space. By contrast, since original ensembles have steeper spectra, they are more structured in phase space.

\subsubsection*{Convergence of ensemble subspace}

In addition to the eigenvalues, the other characteristic of ensembles used in this work is the subspace they span. Even though there are not enough members to estimate the eigenvalues of the true covariance matrix, one may wonder if these subspaces converge to the subspace defined by the true covariance matrix.

As shown in Sec. \ref{covariance_eigendecomposition_sec}, these subspaces are spanned by the eigenvectors of the covariance matrix and, in order to evaluate the stability of the eigenvectors of the covariance matrix with the number of ensemble members, one could consider the angle between the first eigenvector for the 50-members ensemble and the first eigenvector in a smaller ensemble, the angle between the second eigenvector in the 50-members ensemble and the second eigenvector in the same smaller ensemble, etc. But this could lead to a pessimistic evaluation, for example, if the order of two eigenvectors is exchanged between two ensembles. Principal angles introduced in Sec. \eqref{principal_angles_def_sec} are therefore used in this Section. They should be $0$ for the largest ensembles if the ensemble subspaces converge.

The principal angles are computed between the ensemble subspace of the 50-members ensemble and the ensemble subspace of the smaller ensembles extracted from it. However, the subspaces of smaller ensembles are included in the subspace of the 50-members ensemble. Therefore, the principal angles between the subspaces of smaller ensembles and the subspace of the 50-members ensemble are all exactly zero. Because of this, the principal angles are instead computed between the subspace spanned by the 4 first eigenvectors of each smaller ensemble and the subspace spanned by the 4 first eigenvectors of the 50-members ensemble. The first eigenvectors are chosen because they should be the ones that are best approximated, as the relative gaps between eigenvalues are the largest \cite{North1982} (see the Supplementary material for the spectrum of eigenvalues). Note that the first principal angle computed in this way is not necessarily the angle between the first eigenvector in the 50-members ensemble and the first eigenvector in the smaller ensemble.

\begin{figure}
\centering
\hspace*{-2cm}
\includegraphics[scale=0.55]{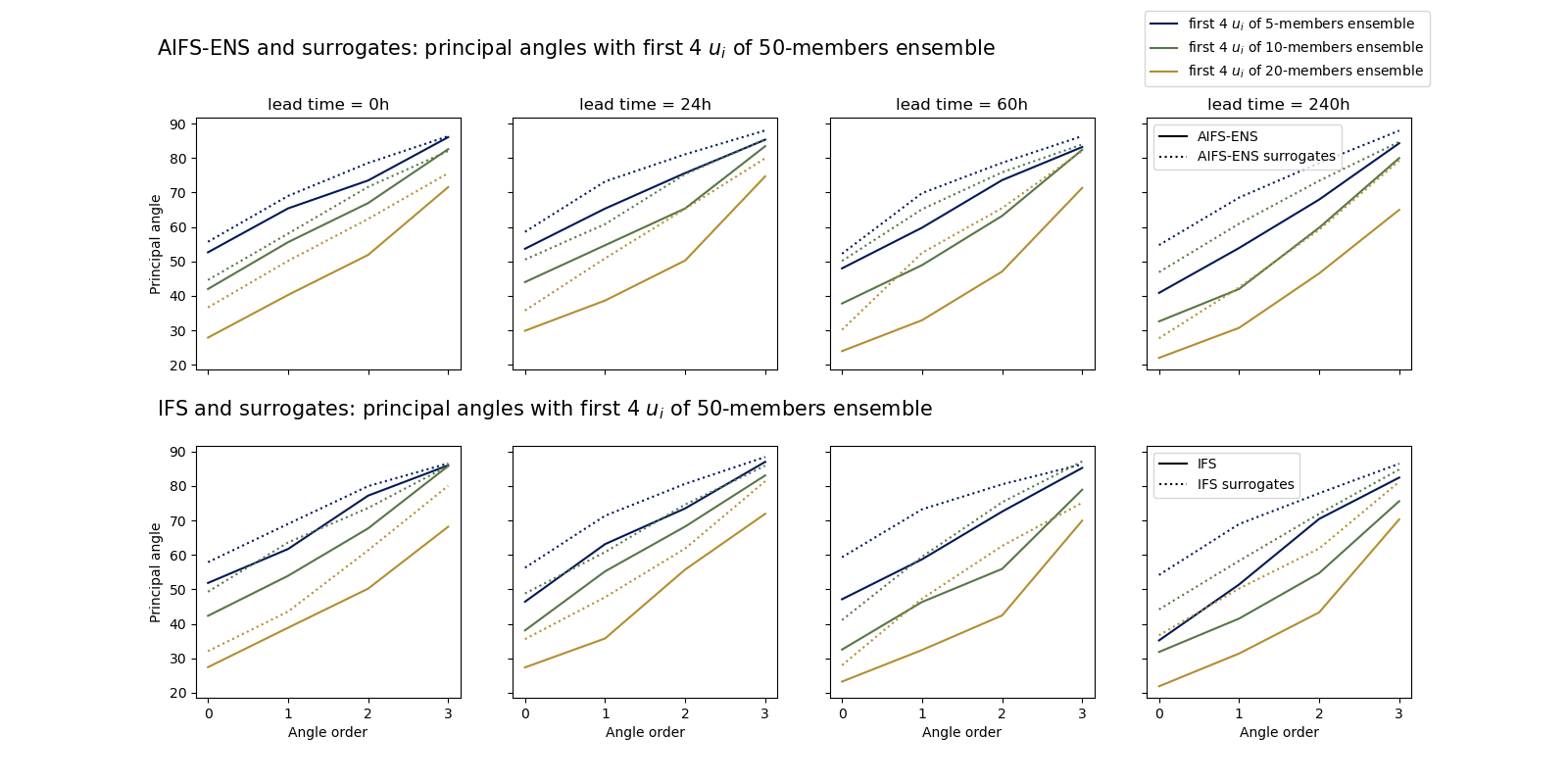}
\caption{Eigenvectors convergence. Shown are the principal angles between the first 4 eigenvectors of the smaller ensembles and the first 4 eigenvectors of the 50-members ensemble. Top row: for AIFS-ENS and AIFS-ENS surrogates, bottom row: for IFS and IFS surrogates (bottom row), for different lead times.}
\label{eigenvectors_convergence_fig}
\end{figure}

Figure \ref{eigenvectors_convergence_fig} shows the principal angles for different lead times, for AIFS-ENS and IFS and their surrogates. The angles are averaged over 100 forecasts. The angles decrease with the number of members in the smaller ensembles, but they do not yet seem to converge, again showing that the covariance matrices are not those of the distributions generating the ensembles. 

The first eigenvectors of surrogate ensembles are random, subject to the power-spectrum constraint on the surrogate $v_i$. The values of the principal angles for surrogate ensembles show that a large part of the principal angle values in AIFS-ENS and IFS ensembles is due to the power spectrum of the $v_i$.

\subsection{Projection on ensemble subspace}

Even though the ensembles are not large enough to accurately define the covariance matrix and its eigenvalues and eigenvectors, one may wonder whether the directions in which these ensembles are aligned are indeed relevant to represent the error. This is measured by the projection of the error on the ensemble subspace given by $\cos\gamma(e, p)$.

\begin{figure}
\centering
\hspace*{-1.5cm}
\includegraphics[scale=0.8]{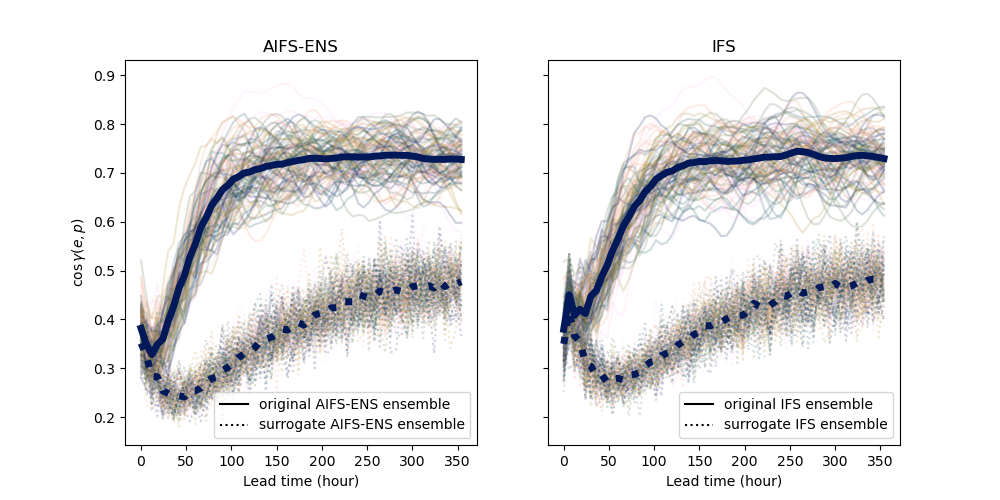}
\caption{$\cos\gamma(e, p)$ for AIFS-ENS (left) and IFS (right) ensembles. Solid lines show $\cos\gamma(e, p)$ for the original ensembles, dotted lines for the surrogate ensembles, thin lines for individual cases, and thick lines for the averages over cases.}
\label{cos_gamma_fig}
\end{figure}

Figure \ref{cos_gamma_fig} displays $\cos\gamma(e, p)$ for the AIFS-ENS and the IFS ensembles, and for their surrogates. As could be expected, these curves are qualitatively very similar to those of \cite{Wei2003}. The difference is that the values of $\cos\gamma(e, p)$ are now higher, most likely due to improved forecasting systems.

For the AIFS-ENS ensembles, $\cos\gamma(e, p)$ consistently starts by decreasing, as if AIFS-ENS was first degrading its prediction. For IFS ensembles, $\cos\gamma(e, p)$ starts by oscillating for the first 24 hours. Original AIFS-ENS and IFS ensembles otherwise behave very similarly: for lead times larger than $\sim 24$ hours, $\cos\gamma(e, p)$ steadily increases and then plateaus around $\sim 0.7$, which reflects the fact that the scale of the error and the scale of the $v_i$ increase and it becomes more and more likely that the error will project on the ensemble subspace.

On the other hand, after a decrease for lead times smaller than 50 hours, the value of $\cos\gamma(e, p)$ for surrogates saturates to $\sim 0.45$. This is because a large part of $\cos\gamma(e, p)$ is due to the shape of the power spectrum of the $v_i$ only. Again, the fact that $\cos\gamma(e, p)$ is significantly non-zero for surrogates shows that a large part of its values is due to the shape of the power spectrum of $v_i$. Conversely, the difference between $\cos\gamma(e, p)$ for original AIFS-ENS and IFS ensembles and for the surrogate ensembles quantifies the skill of the original ensembles to localize the error at long lead times. There is, in particular, almost no skill of the original ensembles at the beginning of the forecast, since $\cos\gamma(e, p)$ is the same as for the surrogate ensembles: $\cos\gamma(e, p)$ equals $0.38$ for the original ensembles and $0.35$ for the surrogates.

One could not expect $\cos\gamma(e, p)$ to be $1$ on average since the ensembles do not have enough members to estimate the covariance matrix and thus to estimate all the directions of instability, unless the ensembles systematically had false directions of instability. For short lead times, many more members are still needed to fully represent all the unstable directions. On the other hand, the fact that $\cos\gamma(e, p)$ is systematically around 0.7 for long lead times hints at the fact that the ensemble size might be close to the ensemble size needed at that lead time to represent all the instabilities.

\subsection{Projection on eigenvectors and relative variance}

\label{projection_p_u_i_relative_variances_results_sec}

\begin{figure}
\centering
\hspace*{-0.5cm}
\includegraphics[scale=0.5]{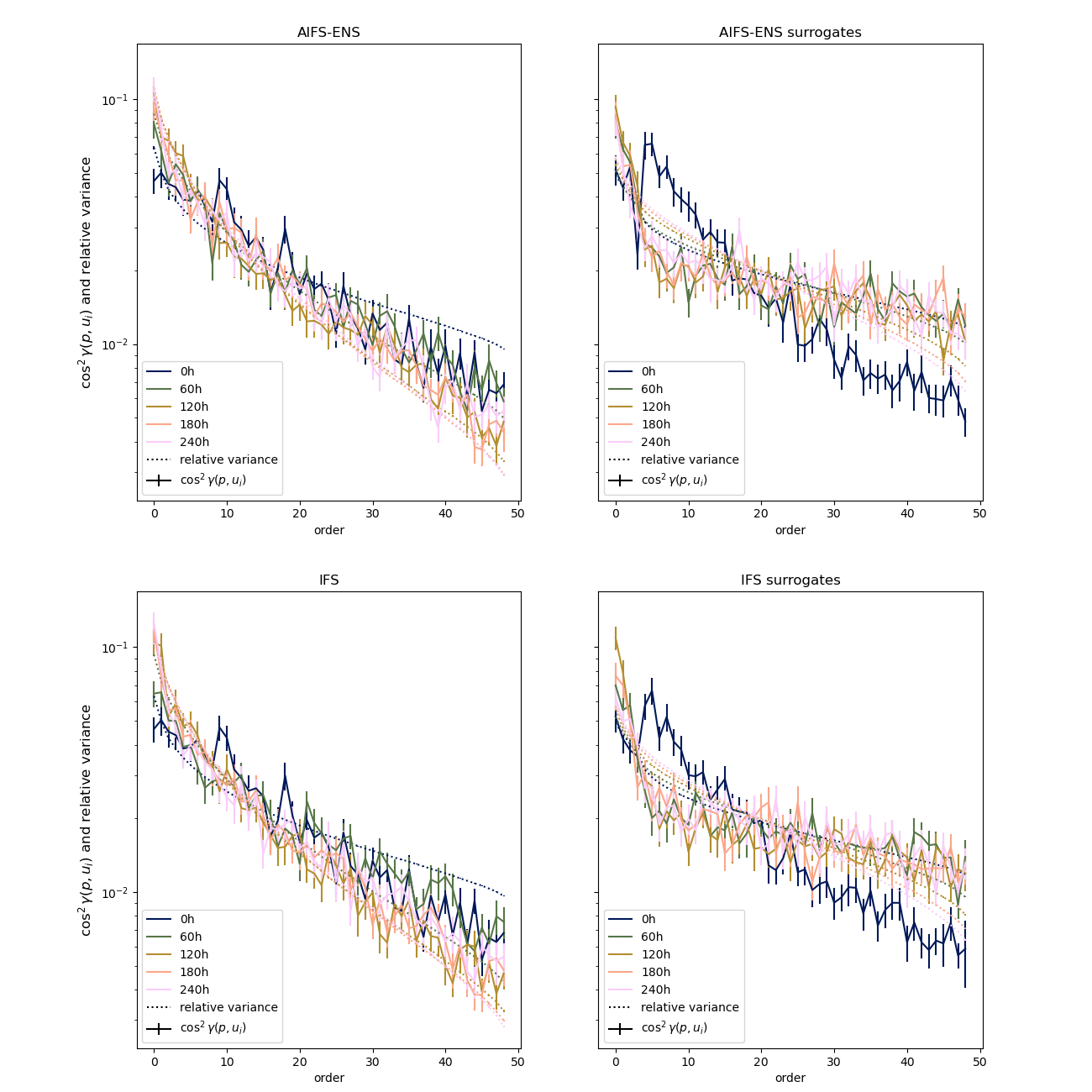}
\caption{Comparison of the projection of the error on the $u_i$ with the relative variances in the ensembles, for AIFS-ENS (top left), AIFS-ENS surrogates (top right), IFS (bottom left) and IFS surrogates (bottom right). The solid lines represent the mean of $\cos\gamma(p, u_i)$ as a function of the order and the dotted lines represent the relative variance of the ensemble at that order. Error bars on solid lines show the standard error of the mean of $\cos\gamma(p, u_i)$, while the standard error of the mean of the relative variances is too small to be displayed. The color represents the lead time. If the ensemble is calibrated for some lead time, the dotted line is within the error bars of the solid line for that lead time.}
\label{cos_gamma_p_u_i_fig}
\end{figure}

As explained in Sec. \ref{projection_p_u_i_relative_variances_sec}, the projection of the error on the eigenvectors, measured by $\cos\gamma(p, u_i)$, should be in average equal to the relative variances in the ensemble. The results of the test of this equality are presented in Fig. \ref{cos_gamma_p_u_i_fig}, again for AIFS-ENS, IFS and their surrogates. The solid lines represent $\cos\gamma(p, u_i)$ and the dotted lines represent the relative variances, each averaged over 100 forecasts. Error bars on solid lines represent the standard error of the mean of $\cos\gamma(p, u_i)$, while the error bars for the relative variances are too small to be displayed. Indeed, the distributions of relative variances are much more compact than those of $\cos\gamma(p, u_i)$: an example of the distribution of $\cos^2\gamma(p, u_i)$ and of the distribution of relative variance for the 100 forecasts used in this work is provided in the Supplementary material. 

If an ensemble is calibrated in all directions for a given lead time, the dotted line for that lead time is within the error bars of the corresponding line for $\cos\gamma(p, u_i)$. If the solid line is higher than the dotted line for some order, the ensemble is overdispersive for that order and for the corresponding lead time. By construction, $\sum_i\cos^2\gamma(p, u_i) = 1$ and the relative variances also sum to 1, so a too large relative variance for some order means that the relative variance is too small for another order.

From Fig. \ref{cos_gamma_p_u_i_fig}, it appears that the relative variances of AIFS-ENS and IFS ensembles generally match quite well the projection of the error on the eigenvectors. At the beginning of the forecast, the relative variances corresponding to eigenvectors whose order is between 5 and 15 or above 25 are too small, while the relative variances for the orders 9, 10, and 18 are too large.

In contrast, the surrogates show a large mismatch at the beginning of the forecast. After that, the relative variances are too small for orders below 12 and too large for other orders.

\subsection{Angles}

\label{angles_sec}

The principal angles introduced in Sec. \ref{principal_angles_def_sec} were computed between the AIFS-ENS and IFS ensemble subspaces to compare how they extend in phase space. The angles between corresponding members in the ensembles were also computed: this is meaningful since the ensembles are initialized with the same perturbations of the initial condition, and this kind of angle describes how an AIFS-ENS member and the corresponding IFS member evolve with respect to each other. These angles capture both model differences and differences in the realized stochastic physics across members.

\begin{figure}
\centering
\hspace*{-1cm}
\includegraphics[scale=0.6]{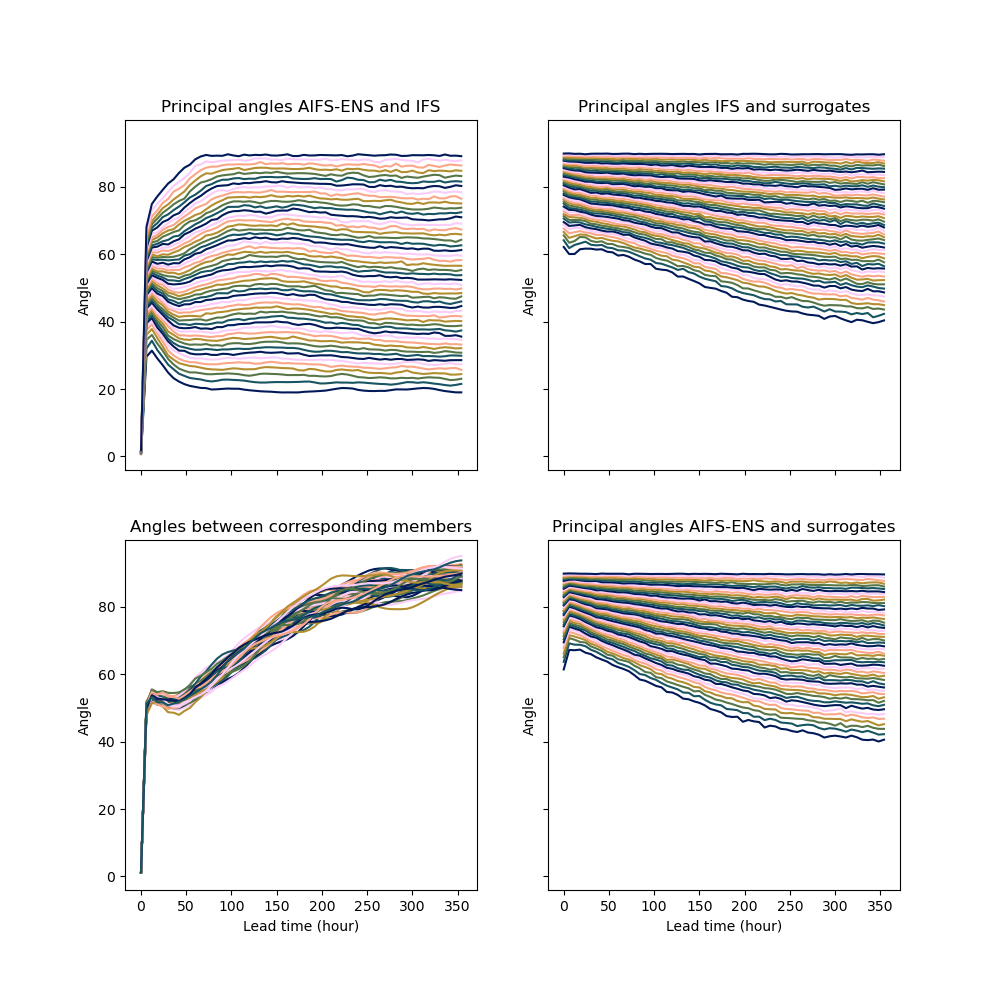}
\caption{Angles between ensembles, averaged over 100 forecasts. Top left: principal angles between AIFS-ENS and IFS ensembles. Top right: principal angles between IFS ensembles and their surrogates. Bottom left: member-by-member angles. Bottom right: principal angles between AIFS-ENS ensembles and their surrogates.}
\label{angles_fig}
\end{figure}

Figure \ref{angles_fig} displays the two different kinds of angles between an AIFS-ENS ensemble, the corresponding IFS ensemble, and their surrogates. Each curve in the top-left panel shows the value of one principal angle between the AIFS-ENS and IFS ensembles, averaged over 100 forecasts. These are zero for the first lead time since both ensembles are initialized with the same perturbations. The principal angles between the two subspaces increase immediately, reaching their final values after 50 hours. These final values are uniformly spread between $20^\circ$ and $90^\circ$, so the ensembles agree almost completely in some directions but not in others. There is no zero principal angle, which means that there is no direction on which the ensembles agree perfectly.

The two right panels show principal angles between IFS and AIFS-ENS ensembles and their respective surrogates. Each curve again represents the value of one principal angle averaged over 100 forecasts. These angles are large at the beginning of the forecast but, as a larger fraction of the error spreads to large scales, the ensembles become more likely to intersect, so the principal angles decrease. However, these principal angles (between AIFS-ENS and IFS and their respective surrogates) are larger than the ones between AIFS-ENS and IFS ensembles, and this indicates that there is more agreement between the AIFS-ENS and IFS ensemble subspaces than between the original ensemble subspaces and their surrogate ensemble subspaces. This shows that, even though the original ensemble subspaces do not overlap, they are more similar to each other than to surrogate ensemble subspaces.

The bottom-left panel displays angles between members initialized with the same perturbations, with one curve for each angle. All angles behave similarly: they increase quickly to $\sim 50$ degrees, plateau there until the lead time reaches 50 hours, and then steadily increase toward 90 degrees. This shows that corresponding AIFS-ENS and IFS members do not evolve in the same way. These angles differ substantially from the principal angles, indicating that the smallest principal angles are between directions defined by linear combinations of several ensemble members.

\subsection{Location of error and eigenvectors in space}

\begin{figure}
\centering
\hspace*{-2.5cm}
\includegraphics[scale=0.6]{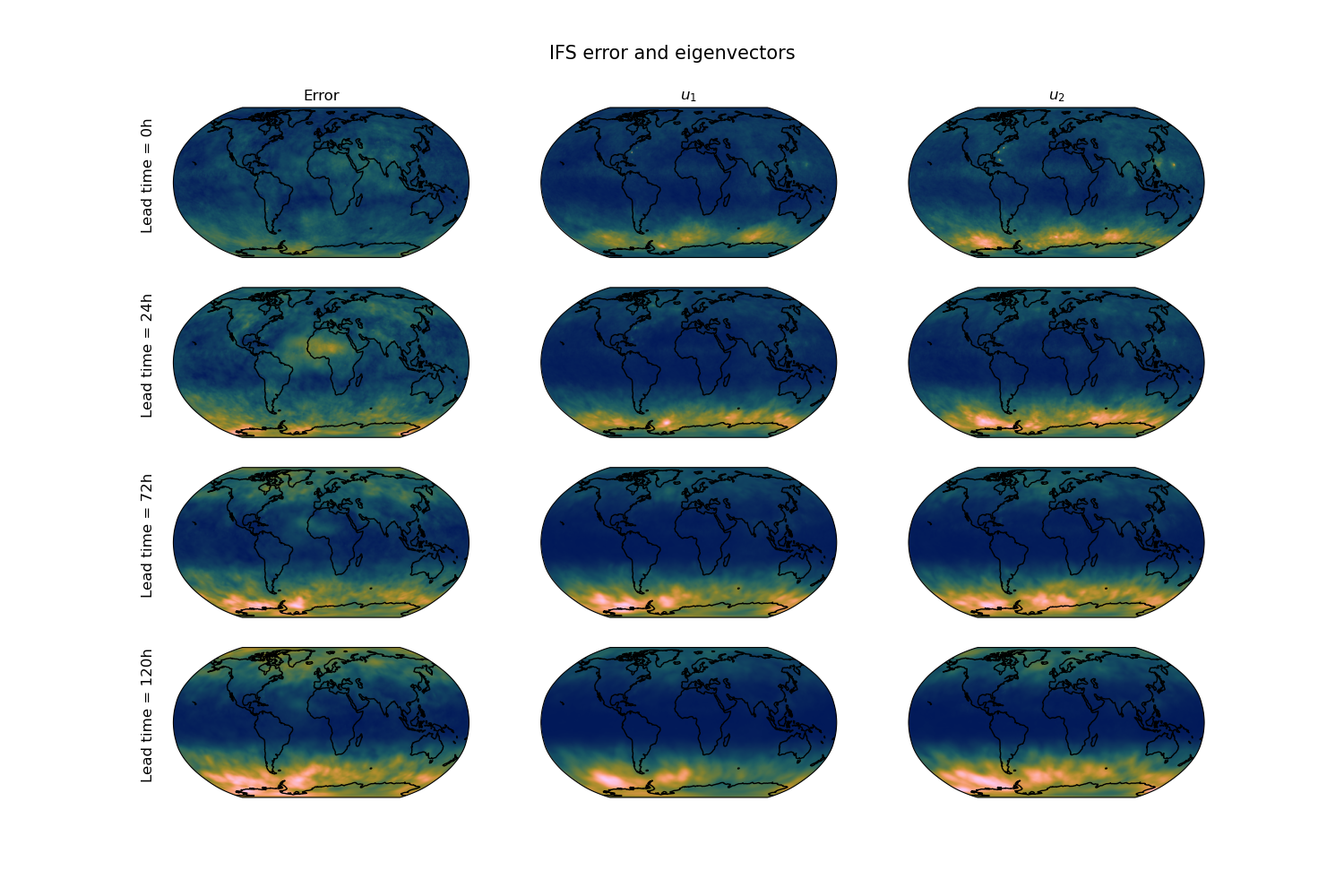}
\caption{Absolute error and eigenvectors averaged over the 51 IFS forecasts of July and August 2025 (austral winter). Each row corresponds to a different lead time, while the first column displays the error $e$, and the second and third columns display the first and second eigenvectors.}
\label{IFS_error_eigenvector_fig}
\end{figure}

\begin{figure}
\centering
\hspace*{-2.5cm}
\includegraphics[scale=0.6]{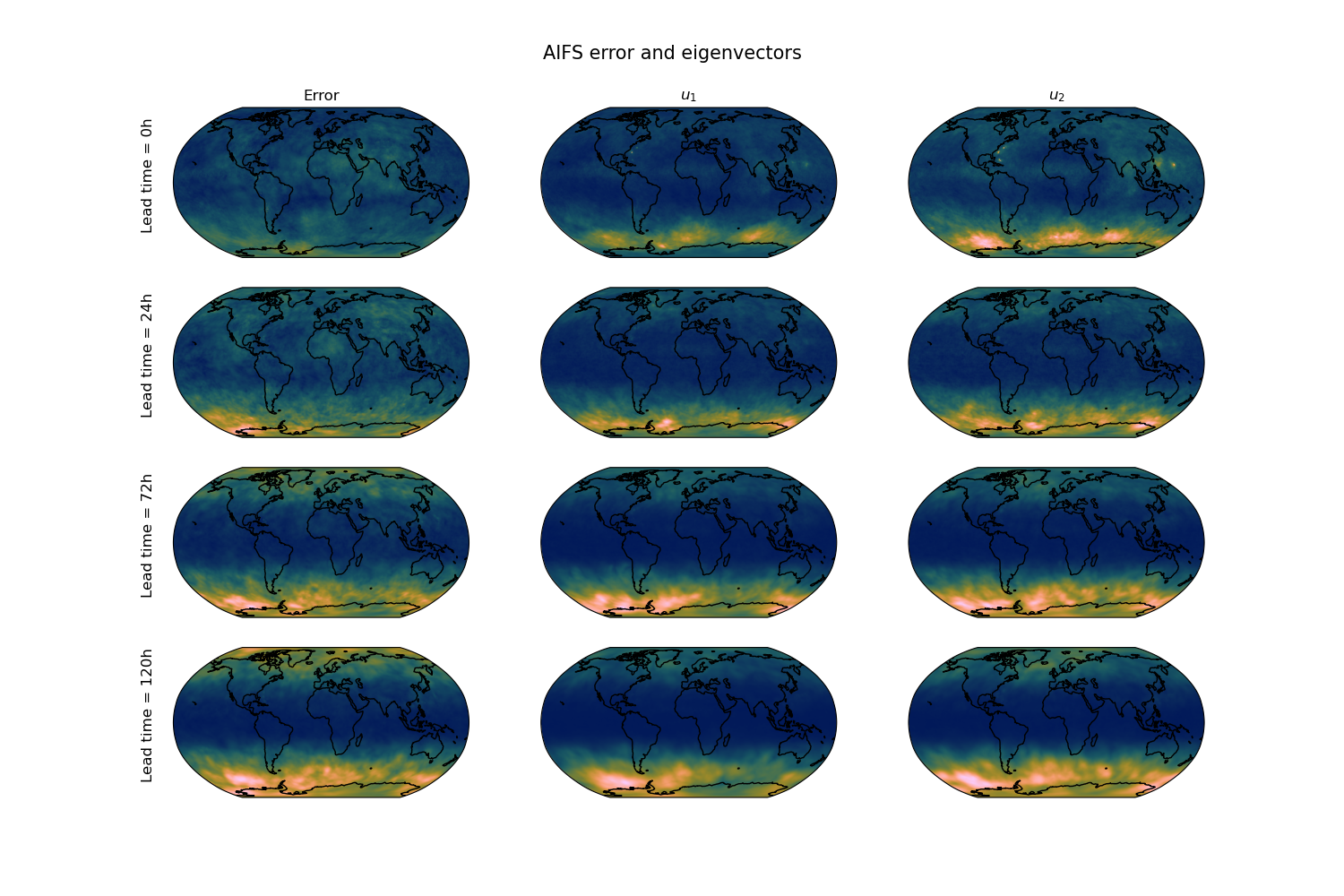}
\caption{Absolute error and eigenvectors averaged over the 51 AIFS-ENS forecasts of July and August 2025 (austral winter). Each row corresponds to a different lead time, while the first column displays the error $e$, and the second and third columns display the first and second eigenvectors.}
\label{AIFS_error_eigenvector_fig}
\end{figure}

Finally, the average location of the error and of the eigenvectors in space is explored in order to support the interpretation of the eigenvectors as modes of uncertainty. Figure \ref{IFS_error_eigenvector_fig} shows the absolute values of the error $e$ and of the first two eigenvectors of IFS ensembles, averaged over forecasts. These quantities are averaged over July and August 2025 forecasts (51 forecasts). Figure \ref{AIFS_error_eigenvector_fig} displays the corresponding quantities for AIFS. In both cases, those errors are visually developing at large synoptic scales, as already revealed previously \citep[e.g.][]{Vannitsem1998,Corazza2003,Feng2018}, and in line with the dominant spectral amplification at the large-scale modes presented earlier.

The error shows no clear systematic pattern at the beginning of the forecasts. At longer lead times, the error mainly develops in the Southern Hemisphere midlatitudes, in the Pacific and Atlantic oceans. The first and second eigenvectors have patterns similar to the error, confirming that they can be interpreted as the main uncertainty modes in the ensembles. The mean absolute values of the eigenvectors are very similar even though there is no zero principal angle between AIFS-ENS and IFS ensemble subspaces. One reason for this could be that the underlying distributions of the ensembles are the same for the two models, but the actual realizations of the ensembles do not overlap.

\section{Discussion and conclusion}

\label{discussion_conclusion_sec}

The error representation in IFS and AIFS-ENS ensembles has been analyzed from a dynamical systems perspective, interpreting the ensembles through the linear subspace spanned by the $v_i$ and the eigenvalues of the ensemble covariance matrix. In contrast to ensemble evaluations that apply a uni-dimensional metric at each grid point, this method evaluates and interprets ensembles as objects in phase space, accounting for spatial patterns among them. The EOFs have a natural role in this context, and the results in Sec. \ref{projection_p_u_i_relative_variances_results_sec} show that they can be used to identify the directions where the largest error is expected.

The evaluation presented in Sec. \ref{projection_p_u_i_relative_variances_results_sec} about the calibration of the ensemble in all directions is very close to what was done in \cite{Leutbecher2014} for singular vectors, in terms of methodology and in terms of results. The fact that both the $u_i$ and the singular vectors can be used to meaningfully decompose the ensemble variance actually suggests that there is a connection between these two sets of vectors. In fact, \cite{Leutbecher2008} argue that the $u_i$ are the future singular vectors (i.e. the singular vectors at the end of their optimization interval) because the covariance matrix evolves with the tangent-linear propagator used to compute the singular vectors. Another equivalent way to understand this connection comes from the fact that initial perturbations align with singular vectors during the forecast, so the ensemble variance mostly increases in the directions of singular vectors. There are, however, differences between the $u_i$ and singular vectors. First, the latter are computed with respect to a model trajectory, while the former are defined with respect to the ensemble mean. Second, singular vectors are usually computed with respect to the energy norm, not with the Euclidean norm of the geopotential as here.

Even though there are too few members in the ensembles to confidently estimate the eigenvalues and the $u_i$ (and thus the ensemble subspaces), the error has a significant projection onto the ensemble subspace. The comparison with $\cos\gamma(e, p)$ for surrogates shows that AIFS-ENS and IFS ensembles have some skill in estimating the orientation of the error in phase space, except at the beginning of the forecast, where the ensembles have surprisingly almost no skill. As pointed out in \cite{Leutbecher2014}, this initial skill would probably be higher if evaluated with the norm used in the data assimilation scheme.

When the total norm of the error and the total variance of the ensembles are rescaled to match each other, the ensembles correctly estimate the variance of the error in each direction. The analysis with principal angles between AIFS-ENS and IFS ensembles reveals that the subspaces do not overlap, but they are more similar to each other than to the surrogate ensemble subspaces. This suggests that combining both ensembles into a 100-members ensemble could bear some useful information in operational forecasting.

Interestingly, AIFS-ENS ensembles behave very similarly to IFS ensembles even though their ensemble subspaces differ. In particular, AIFS-ENS can estimate the phase-space direction of the error of its ensemble mean. This is remarkable since the ensembles are initialized partly from singular vectors computed with IFS, not with AIFS-CRPS. Therefore, one could argue that AIFS-CRPS learned to represent real atmospheric instabilities in its ensembles. But it is difficult to assess the skill of AIFS-CRPS itself from the present work precisely because AIFS-ENS is initialized with IFS initial conditions.

Future work could take several directions. First, a major problem for the covariance matrix in global forecasts is that 50 ensemble members cannot be used to estimate it accurately. In other words, the covariance matrix is rank deficient. This likely means the number of uncertainty modes in the underlying distribution generating the members exceeds 50. In this case, the ensembles should have more than 50 members to fully sample this distribution. A similar problem arises in data assimilation \cite{Carrassi2018}, and one mitigation is to work in subdomains where ensemble members can be treated as independent because correlation often decays exponentially with distance. This amounts to splitting the phase space into lower-dimensional spaces. Another option is to adapt the tools to be scale-dependent; the advantage is that it would make explicit that different scales dominate the scalar product in Eq. \eqref{scalar_product_SH_eq}.

The framework adopted in this work is completely linear because linear combinations of the $v_i$ are needed to define the $u_i$ and the eigenvalues. But linear combinations of ensemble members become inconsistent in the nonlinear regime, which cannot be neglected at the end of the forecast. For example, CNOPs are designed to capture nonlinear effects that singular vectors cannot represent, so a linear combination of Conditional Nonlinear Optimal Perturbations (CNOPs, \cite{Mu2003}) does not evolve into the corresponding linear combination of the evolved CNOPs. The interpretation of such a linear combination is therefore unclear. For this reason, the PECA score (which is analogous to $\cos\gamma(e, p)$) used in \cite{Duan2022, Ji2025} and \cite{Ma2025} does not capture the nonlinearities described by the CNOPs. It would thus be valuable to evaluate and interpret these perturbations with phase-space techniques that capture nonlinearities.

It might also be interesting to understand why $\cos\gamma(e, p)$ decreases at short lead times for AIFS-ENS, to gain additional insight into the dynamics learned by the model. This might be related to the decrease in $\cos\gamma(e, p)$ for surrogate ensembles. Extensive spatial visualizations of the error and of the eigenvectors could also help to understand this decrease in $\cos\gamma(e, p)$.

\codedataavailability{The code used in this work is available at the Zenodo repository \url{https://doi.org/10.5281/zenodo.18341086}.}

\appendix
\section{Equivalence of $\cos\gamma(e, p)$ and PECA}

\label{cos_gamma_PECA_sec}

The cosine of the angle between the error and perturbation vectors is the Perturbation versus Error Correlation Analysis (PECA) score \citep{Wei2003}, where the control forecast is replaced by the ensemble mean. Indeed, the Pattern Correlation Anomaly (PAC) used in PECA is the formula for the cosine of an angle. In PECA, this angle is the one between the error and the vector $P$ that minimizes
\begin{align}
    L(\alpha_i) = \left|e - P\right|^2, \text{ with } P = \sum_i\alpha_iv_i.
\end{align}
The vector $P$ is a vector in the space spanned by the $v_i$, so that the solution of this minimization problem is quite simple: at the minimum, $P$ is the projection of $e$ on this space. This can be seen easily since $P$ can be expressed in terms of the eigenvectors as $P = \sum_j\beta_ju_j$ with $\beta_j = \sum_i\alpha_ic_{ij}$, and the function to minimize is now
\begin{align}
    L(\beta) = \left|e - \sum_j\beta_ju_j\right|^2,
\end{align}
The minimum is obtained by imposing $\frac{\partial L}{\partial \beta_j} = 0$:
\begin{align}
    \left(e - \sum_k\beta_ku_k\right) \cdot u_j = 0 \Leftrightarrow e\cdot u_j - \sum_k\beta_ku_k\cdot u_j = 0 \Leftrightarrow \beta_j = e\cdot u_j
\end{align}
so that $P$ is indeed the projection of $e$ on the ensemble subspace: $P = p$.

The PECA is thus the angle between the error of the control run and $p$, and it is equal to $\cos\gamma(e, p)$ if the control run is replaced by the ensemble mean for both the definition of the error and the definition of the $u_i$.

\noappendix

\authorcontribution{M.B. and S.V. designed the study. M.B. wrote the code and performed the computations. M.B. and S.V. interpreted the results and wrote the manuscript.}

\competinginterests{The authors have no competing interests.}

\begin{acknowledgements}
The authors warmly thank Zied Ben Bouallègue for discussions in the early stages of the work and for comments on the manuscript. This research has been supported by the Belgian Federal Science Policy Office (BELSPO) under contract number P4S/251/ForeBAt.
\end{acknowledgements}

\bibliographystyle{copernicus}
\bibliography{biblio.bib}

\end{document}